\documentclass[twoside,11pt]{article}
\usepackage{a4}
\usepackage{bm}
\usepackage{amsmath,bbm,fancyhdr,epsfig,umlaut,amssymb,psfrag,multirow,exscale,caption,graphicx,slashbox,color}
 \allowdisplaybreaks[1] 
\numberwithin{equation}{section}
\newcommand{\be}{\begin{equation}}
\newcommand{\ee}{\end{equation}}
\newcommand{\bea}{\begin{eqnarray}}
\newcommand{\eea}{\end{eqnarray}}
\newcommand{\ba}{\begin{align}}
\newcommand{\ea}{\end{align}}
\newcommand{\refe}[1]{Eq.\/~(\ref{#1})}
\newcommand{\refes}[2]{Eqs.\/~(\ref{#1}--\ref{#2})}
\newcommand{\refec}[2]{Eqs.\/~(\ref{#1},\:\ref{#2})}
\newcommand{\transpose}{\mathrm{T}}

\newcommand{\R}{\mathbbm{R}}

\newcommand{\uosp}{U\!O\!Sp(1|2)}

\definecolor{grey}{rgb}{0.45,0.45,0.45}

\title{
\vskip -70pt
\begin{flushright}
{\normalsize \ DAMTP-2005-9}\\
\end{flushright}
\vskip 25pt {\bf Graded Majorana spinors} } \vspace{1.4cm} {\makeatletter
\author{A.~F.~Kleppe\thanks{e-mail address:
    A.F.Kleppe@damtp.cam.ac.uk}\hspace{5pt}\thanks{A.F.K. previously published under the name A.~F.~Schunck}
\hspace{1pt} and Chris Wainwright{\thanks{e-mail address:
C.J.Wainwright@damtp.cam.ac.uk}} \\ \small{\textsl{Department of Applied
    Mathematics and Theoretical Physics}}
\\ \small{\textsl{University of Cambridge}} \\
\small{\textsl{Wilberforce Road, Cambridge CB3 0WA, England}}}
\makeatother}
\date{\today}
\begin{document}
\maketitle
\begin{abstract}
In many mathematical and physical contexts spinors are treated as
Grassmann odd valued fields. We show that it is possible to extend the
classification of reality conditions on such spinors by a new type of
Majorana condition. In order to define this \emph{graded Majorana}
condition we make use of pseudo-conjugation, a rather unfamiliar extension
of complex conjugation to supernumbers. Like the symplectic Majorana
condition, the graded Majorana condition may be imposed, for example, in
spacetimes in which the standard Majorana condition is inconsistent.
However, in contrast to the symplectic condition, which requires
duplicating the number of spinor fields, the graded condition can be
imposed on a \emph{single} Dirac spinor. We illustrate how graded Majorana
spinors can be applied to supersymmetry by constructing a globally
supersymmetric field theory in three-dimensional Euclidean space, an
example of a spacetime where standard Majorana spinors do not exist.
\end{abstract}
\section{Introduction}
One of the key ingredients to a deep understanding of the mathematical
concept of spinor fields has been the complete classification of all
possible types of reality conditions that can be imposed on spinors in a
given spacetime. If spinors are treated as ordinary fields, this
classification of possible reality conditions, normally referred to as
\emph{Majorana} conditions, has been given in \cite{townsend_kugo}.
However, though this classification of Majorana conditions nicely extends
to spinors treated as Grassmann odd valued fields, as is the case for
example in supersymmetric theories, it turns out not to be complete. To
see this, note first that the components of such Grassmann odd valued
spinor fields are given by anticommuting supernumbers. Since a Majorana
condition relates a spinor to its complex conjugate, extending the notion
of a Majorana condition to such anticommuting spinor fields implies that
one first has to extend the notion of complex conjugation to supernumbers.
There is, however, an ambiguity in defining this extension, leading to at
least two inequivalent notions of complex conjugation of supernumbers.
These we will refer to as standard complex conjugation \cite{dewitt} and
\textit{pseudo-conjugation} \cite{scheunert}, respectively. While standard
complex conjugation essentially leads to the classification of Majorana
conditions as given in \cite{townsend_kugo}, we show that
pseudo-conjugation makes it possible to define a genuinely new type of
Majorana spinor, which we will refer to as \emph{graded Majorana}.

It should be pointed out that the existence of such reality
conditions in the special case of four-dimensional Euclidean space
has already been discussed in \cite{hawking, manin, ivanov}. In this
paper we will show how this special case is part of the wider and
more general scheme of graded Majorana spinors which, as we shall
see, are entirely complementary to standard
Majorana spinors.

\section{Pseudo-conjugation}\label{sec:pseudo}
Let us first briefly comment on the properties of standard complex
conjugation and pseudo-conjugation, respectively. While the operation of
standard complex conjugation on supernumbers is an involution,
pseudo-conjugation in contrast is a \emph{graded} involution. Denoting the
operation of standard complex conjugation by $*$ and pseudo-conjugation by
$\diamond$ we thus have
\begin{equation}
z^{**}=z, \qquad z^{\diamond\diamond} = (-1)^{\epsilon_z}z
\label{eqn:dbldiamond}.
\end{equation}
Here $\epsilon_z=0$ if $z$ is an even (commuting) supernumber, and
$\epsilon_z=1$ if $z$ is odd (anticommuting). It is this property of
pseudo-conjugation which will enable us later to define a new kind of
Majorana spinor. Additionally, standard complex conjugation and
pseudo-conjugation, respectively, satisfy the properties
\begin{subequations}
\begin{alignat}{2}
(z+w)^* & = z^*+w^*, & \qquad (z+w)^\diamond & = z^\diamond+w^\diamond, \\
(zw)^* & = w^*z^*, & \qquad (zw)^\diamond & = z^\diamond w^\diamond.
\end{alignat}
\end{subequations}
Note that both types of conjugation reduce to ordinary complex conjugation
on ordinary numbers.

A general supernumber can be expanded in the generators $\zeta_i$,
$i=1,\ldots, N$, of a Grassmann algebra as
\be\label{eqn:supernum}
z=z_0+z_i\zeta_i+z_{ij}\zeta_i\zeta_j+z_{ijk}\zeta_i\zeta_j\zeta_k+\ldots
\, .
\ee
Here the coefficients $z_0, z_i, \ldots $ are ordinary complex numbers.
With respect to standard complex conjugation the generators will be taken
to be real, i.e., $\zeta_i^*=\zeta_i$. However, imposing a similar reality
condition on the generators using pseudo-conjugation will be inconsistent
with \refe{eqn:dbldiamond}. Instead, without loss of generality, we will
impose
\be
\zeta_{2i}^\diamond=\zeta_{2i-1}, \qquad
\zeta_{2i-1}^\diamond=-\zeta_{2i}.
\ee
This requires the number $N$ of Grassmann generators to be even or---as
one normally considers in the context of supersymmetric
theories---infinite. Note that $\zeta_i^{*\diamond}=\zeta_i^{\diamond *}$,
from which it follows that standard complex conjugation commutes with
pseudo-conjugation on arbitrary supernumbers.

As we shall see, it will be convenient to split the supernumber $z$ into a
sum of two parts
\begin{subequations}
\begin{alignat}{1}
& \hspace{55pt}z=z_1+z_2, \label{eqn:split} \\ \label{eqn:split2} &
z_1=\frac{1}{2}(z+z^{*\diamond}), \qquad z_2=\frac{1}{2}(z-z^{*\diamond}).
\end{alignat}
\end{subequations} Using this splitting we define an invertible map $f$ on even supernumbers
$z$
\begin{subequations}
\begin{alignat}{2}
\label{eqn:f} f&: z \to \tilde{z} & = z_1 + i z_2,\\
f^{-1}&: \tilde{z} \to z & = \tilde{z}_1 - i \tilde{z}_2,
\end{alignat}
\end{subequations}
with $\tilde{z}_{1,2}$ defined analogously to $z_{1,2}$ in
\refe{eqn:split2}. This map satisfies the property
\begin{equation}
f(z^\diamond) = f(z)^*,
\end{equation}
which can be shown using the fact that $z_1^\diamond = z_1^*$ and
$z_2^\diamond = -z_2^*$. It follows that imposing a pseudo-reality
condition $z = z^\diamond$ on an arbitrary even supernumber $z$ is
equivalent to imposing the standard reality condition $f(z) = f(z)^*$ on
the supernumber $f(z) = \tilde{z}$.

In Section \ref{sec:reality} we will consider how pseudo-conjugation may
be used to impose reality conditions on spinors, the components of which
are taken to be anticommuting supernumbers. However, we first need to
recall some results about Clifford algebras, as discussed in
\cite{townsend_kugo}.

\section{Clifford algebras in $d$-dimensions}
\label{sec:spinors} The Clifford algebra in $d$ spacetime dimensions is
given by
{\renewcommand{\arraystretch}{1.2}%
\begin{equation}
\begin{array}{cc}
\lbrace\gamma^\mu, \gamma^\nu \rbrace = 2 \eta^{\mu \nu}\mathbbm{1}, \\
\eta^{\mu \nu} = \mathrm{diag} (\underbrace{+ + \cdots +}_t \underbrace{-
- \cdots -}_s ),
\end{array}
\end{equation}}%
with $d = t + s$. The $\gamma^{\mu}$ are represented by $2^{\lfloor
d/2\rfloor} \times 2^{\lfloor d/2\rfloor}$ matrices, which may be chosen
such that
\begin{subequations}
\begin{alignat}{2}
\gamma^{\mu\dagger} & = \gamma^\mu, \quad & \mu & = 1, \ldots, t, \\
\gamma^{\mu\dagger} & = - \gamma^\mu, \quad & \mu & = t + 1, \ldots, d.
\end{alignat}
\end{subequations}
Defining $A = \gamma^1 \cdots \gamma^t$ we then have
\begin{equation}
\gamma^{\mu\dagger} = -(-1)^t A \gamma^\mu A^{-1}.
\end{equation}
In even dimensions we can introduce the matrix
\begin{equation}
\Gamma_5 = (-1)^{(t - s)/4} \gamma^1 \cdots \gamma^d,
\end{equation}
which satisfies $(\Gamma_5)^2 = 1$ and is, up to proportionality, the
unique matrix which anticommutes with all $\gamma^\mu$, $\mu = 1, \ldots,
d$. As $\pm \gamma^{\mu*}$ form an equivalent representation of the
Clifford algebra, there exists an invertible matrix $B$ such that
\begin{equation}
\gamma^{\mu *} = \eta B \gamma^\mu B^{-1}, \qquad \eta = \pm 1,
\end{equation}
where $\eta$ can be shown to depend on the signature of the metric, see
Table \ref{spinortable}. Note that in even dimensions, where $t-s$ will
also be even, we always have a choice of $\eta = \pm 1$, whereas in odd
dimensions $\eta$ is fixed. $B$ is unitary and satisfies the condition
\begin{equation}
B^* B = \epsilon \mathbbm{1}, \qquad \epsilon = \pm 1,
\end{equation}
where $\epsilon$ depends on the signature of the metric as well as on the
value of $\eta$ as displayed in Table \ref{spinortable}. Note that $B$ is
only defined up to an overall phase.

The charge conjugation matrix $C$ is defined by
\begin{equation}
\label{eqn:chargeconjugationmatrix} C = B^\transpose A.
\end{equation}
Using the properties of $A$ and $B$ one finds that $C^\dagger C =
\mathbbm{1}$ and
\begin{align}
{\gamma^\mu}^\transpose & = (-1)^{t + 1}\eta C \gamma^\mu C^{-1},\\
C^\transpose & = \epsilon \eta^t (-1)^{t(t -1)/2}C.
\end{align}
The last two equations can be combined to give
\begin{equation}\label{eqn:gammaCinversetr}
(\gamma^\mu C^{-1})^\transpose = (-1)^{t + 1 + t(t - 1)/2} \epsilon
\eta^{t + 1}(\gamma^\mu C^{-1}).
\end{equation}
Additionally we have that
\begin{equation}
\label{eqn:gammaCinverse*} (\gamma^\mu C^{-1})^* = \eta^{t + 1}B
\gamma^\mu C^{-1} B^\transpose.
\end{equation}
These two relations will be important when considering super Poincar\'{e}
algebras in different signatures, see Section \ref{sec:applications}.

In even dimensions, as there is a choice of $\eta = \pm 1$, let us define
$B_\pm$ such that
\begin{equation}
\gamma^{\mu *} = \eta_\pm B_\pm \gamma^\mu B^{-1}_\pm, \qquad B_\pm^*B_\pm
= \epsilon_\pm \mathbbm{1}.
\end{equation}
Here $\eta_\pm = \pm 1$ and $\epsilon_\pm$ is the value of $\epsilon$
corresponding to $\eta_\pm$ in a given signature. Correspondingly we
define $C_\pm=B_\pm^\transpose A$.

Interestingly $B_+$ and $B_-$ are related by
\begin{equation}\label{eqn:relationB+B-}
B_+ = \lambda B_- \Gamma_5,
\end{equation}
where $\lambda$ is an arbitrary phase factor. This relation seems to have
been overlooked in the literature. To prove \refe{eqn:relationB+B-} note
that
\begin{equation}
B_-^{-1}B_+ \gamma^\mu B_+^{-1} B_- = B_-^{-1} \gamma^{\mu *} B_- =
-\gamma^\mu,
\end{equation}
hence $B_-^{-1} B_+$ anticommutes with all the gamma matrices and as such
must be proportional to $\Gamma_5$. Unitarity of both $B_\pm$ and
$\Gamma_5$ restrict $\lambda$ such that $|\lambda|^2 = 1$.

Note that using the relation between $B_+$ and $B_-$,
\refe{eqn:relationB+B-}, we find $\epsilon_+ \mathbbm{1}=  B_+^*B_+ =
|\lambda|^2 (-1)^{(t - s)/2} \epsilon_- \mathbbm{1}$, and hence we see
that
\begin{equation}
\label{eq:relationepsilon+epsilon-} \epsilon_+ = (-1)^{(t - s)/2}
\epsilon_-,
\end{equation}
which is in agreement with Table \ref{spinortable}.
\begin{table}[t!]
\centering {\renewcommand{\arraystretch}{1.2}
\begin{tabular}{|c|c|c|c|c|}
\hline
$\: t-s$ mod $8\:$ & $\: 0,1,2\:$ & $\: 0,6,7\:$ & $\: 4,5,6\:$ & $\: 2,3,4\:$ \\
\hline
$\epsilon$ & $+1$ & $+1$ & $-1$ & $-1$ \\
\hline
$\eta$ & $+1$ & $-1$ & $+1$ & $-1$ \\
\hline \hline
spinor type & $M'$ & $M$ & $gM'$ & $gM$ \\
\hline
\end{tabular}
} \caption{Possible values of $\epsilon$ and $\eta$ in all signatures.}
\label{spinortable}
\end{table}

\section{Reality conditions on spinors}\label{sec:reality}
In many contexts spinors are treated as Grassmann odd valued fields,
i.e.~the $2^{\lfloor d/2\rfloor}$ components of a general Dirac spinor are
given by anticommuting complex supernumbers. Depending on the signature
of the spacetime under consideration such spinors can be constrained by
reality conditions that are both consistent with the Dirac equation and
Lorentz covariant. Reality conditions that satisfy these requirements are
normally referred to as \emph{Majorana} conditions. Conventionally, only
standard complex conjugation of supernumbers has been used to impose such
Majorana conditions. In this section we shall show how, by using
pseudo-conjugation of supernumbers, a genuinely new type of Majorana
condition can be defined.

\subsection{Standard and symplectic Majorana conditions}
Let us first consider signatures in which there exists a matrix $B$ for
which $\epsilon=+1$, i.e.\/~$B^*B = \mathbbm{1}$, see Table
\ref{spinortable}. We may use this matrix $B$ to impose the standard
\emph{Majorana} condition
\begin{equation}\label{eqn:majorana}
\psi=B^{-1}\psi^*.
\end{equation}
Note that imposing such a condition will not be consistent if
$\epsilon=-1$ since $\psi=B^{-1}(B^{-1}\psi^*)^*=(B^*B)^{-1}\psi=\epsilon
\psi$.

In those signatures where there are only matrices $B$ for which
$\epsilon=-1$ one normally introduces a pair (or more generally an even
number) of Dirac spinors $\psi^{(i)}$, $i=1,2$, and imposes the
\emph{symplectic} Majorana condition
\begin{equation}\label{eqn:symplecticmajorana}
\psi^{(i)}=\epsilon^{ij}B^{-1}(\psi^{(j)})^* \quad \mathrm{for} \quad B^*B
= -\mathbbm{1}
\end{equation}
where $\epsilon^{ij}=-\epsilon^{ji}$ with $\epsilon^{12}=+1$. This
condition reduces the degrees of freedom of the pair of spinors down to
that of a single spinor with no reality condition imposed. Therefore,
since a second spinor is initially introduced in order to impose the
symplectic Majorana condition, the number of degrees of freedom is not in
effect reduced.

\subsection{Graded Majorana conditions}
We shall now show that in signatures in which there exists a matrix $B$
for which $\epsilon=-1$, i.e.\/~$B^*B= -\mathbbm{1}$, we can---by making
use of pseudo-conjugation---define an alternative Majorana condition that,
unlike the symplectic one, does not require duplicating the number of
fields, but instead can be imposed on a \emph{single} spinor.
We propose the condition
\begin{equation}\label{eqn:gradedmajorana}
\psi=B^{-1}\psi^\diamond.
\end{equation}
Now, since the components of $\psi$ are anticommuting supernumbers, we
have from \refe{eqn:dbldiamond} that $\psi^{\diamond\diamond}=-\psi$,
hence $\psi = B^{-1}(B^{-1}\psi^\diamond)^\diamond =
(B^*B)^{-1}\psi^{\diamond\diamond} = -\epsilon\psi$ and so the equation is
consistent for $\epsilon=-1$. Note that here we have used $B^\diamond=B^*$
since $B$ is a matrix of ordinary complex numbers. As pseudo-conjugation
is a graded involution we will refer to spinors satisfying
\refe{eqn:gradedmajorana} as \emph{graded Majorana} spinors.

To be complete we also note here that, in those signatures for which there exists a
matrix $B$ for which $\epsilon=+1$, pseudo-conjugation may be used to
define a \emph{graded} symplectic Majorana condition
\begin{equation}\label{eqn:gradedsymplecticmajorana}
\psi^{(i)}=\epsilon^{ij}B^{-1}(\psi^{(j)})^\diamond \quad \mathrm{for} \quad B^*B= +\mathbbm{1}.
\end{equation}
\refec{eqn:gradedmajorana}{eqn:gradedsymplecticmajorana} thus
constitute the graded counterparts of
\refec{eqn:majorana}{eqn:symplecticmajorana} and highlight how graded Majorana
conditions should be treated on an equal footing with the standard
Majorana conditions.

In the next section we will show how reality conditions using standard
complex conjugation and pseudo-conjugation, respectively, can be thought
of as equivalent in terms of the number of constraints they impose on a
spinor.

\subsection{Equivalence of reality conditions}
\label{sec:equivalence_reality_conditions} Just as the standard
Majorana condition of \refe{eqn:majorana} is covariant under Lorentz
transformations so, too, is the graded Majorana condition of
\refe{eqn:gradedmajorana}. For the purpose of analyzing the number
of constraints, however, we shall also consider more general reality
conditions that may not necessarily be so. Let us introduce
$2^{\lfloor d/2\rfloor} \times 2^{\lfloor d/2\rfloor}$ matrices $M$
and $N$ satisfying $M^*M = +\mathbbm{1}$ and $N^*N=-\mathbbm{1}$,
respectively (where we require $d > 1$ for the matrix $N$ to exist).
Then consider reality conditions of the form $\psi=M^{-1}\psi^*$ and
$\psi=N^{-1}\psi^\diamond$, encompassing the standard and graded
Majorana conditions, respectively. In particular these conditions
shall be replaced with the corresponding Majorana conditions,
\refec{eqn:majorana}{eqn:gradedmajorana}, as long as the appropriate
matrices $B$ exist.

In order to show that the number of constraints imposed on a spinor is the
same for both $\psi=M^{-1}\psi^*$ and $\psi=N^{-1}\psi^\diamond$ we will
use an argument analogous to that for an even supernumber as discussed in
Section \ref{sec:pseudo}. Consider the split of
\refec{eqn:split}{eqn:split2} applied to each of the components of the
spinor $\psi$, resulting in
\begin{subequations}
\begin{alignat}{1}
& \hspace{55pt}\psi=\psi_1+\psi_2, \label{eqn:splitpsi}\\
\label{eqn:splitpsi2} & \psi_1=\frac{1}{2}(\psi+\psi^{*\diamond}), \qquad
\psi_2=\frac{1}{2}(\psi-\psi^{*\diamond}).
\end{alignat}
\end{subequations}
Using the fact that $\psi_1^*=\psi_2^\diamond$ and
$\psi_2^*=-\psi_1^\diamond$ it is easily seen that the following two
equivalences hold
\begin{equation}
\label{eqn:standard-symplectic} \psi=M^{-1}\psi^* \iff
\left\{\begin{array}{l}
\psi_1= M^{-1}\psi_2^\diamond \\
\psi_2= - M^{-1}\psi_1^\diamond
\end{array}\right.
\end{equation}
and
\begin{equation}
\label{eqn:graded-symplectic} \psi=N^{-1}\psi^\diamond \iff
\left\{\begin{array}{l}
\psi_1=-N^{-1}\psi_2^* \\
\psi_2=N^{-1}\psi_1^*
\end{array}\right.
\end{equation}
In those signatures where there exists a matrix $B$ such that $B^*B =
-\mathbbm{1}$, \refe{eqn:graded-symplectic} shows how a graded Majorana
condition imposed on the spinor $\psi$ can be restated as a symplectic
Majorana condition imposed on the \emph{split} fields $\psi_{1,2}$ of
\refe{eqn:splitpsi}. Note, however, that the symplectic Majorana condition
is being imposed on the \emph{internal} supernumber structure of a
\emph{single} spinor. Conversely, in those signatures where there exists a
matrix $B$ such that $B^*B = \mathbbm{1}$, we see from
\refe{eqn:standard-symplectic} that the standard Majorana condition is
equivalent to a graded symplectic Majorana condition being imposed on
the split fields $\psi_{1,2}$. Also in this case the symplectic condition
is imposed on the internal supernumber structure of a single spinor.

Let us now define the quantity
\begin{equation}
\label{eqn:psitilde} \tilde{\psi}=\mu^*\psi_1+\mu M^* N \psi_2
\end{equation}
where $\mu$ is some non-zero, ordinary complex constant. The relationship
of \refe{eqn:psitilde} may be inverted to give $\psi$ in terms of
$\tilde{\psi}$. To see this note that if we split $\tilde{\psi}$ as in
\refec{eqn:splitpsi}{eqn:splitpsi2} we have
\begin{subequations}
\begin{align}\label{eqn:psitilde1ofpsi}
\tilde{\psi}_1 &= \frac{1}{2} \big(( \mu^*+\mu M^* N)\psi_1-(\mu^*-\mu M^*
N)\psi_2 \big), \\ \label{eqn:psitilde2ofpsi}\tilde{\psi}_2 &= \frac{1}{2}
\big(( \mu^*-\mu M^* N)\psi_1+(\mu^*+\mu M^* N)\psi_2 \big),
\end{align}
where we have used that $\psi_1^*=\psi_2^\diamond$ and
$\psi_2^*=-\psi_1^\diamond$.
\end{subequations}
We then find
\begin{subequations}
\begin{align}\label{eqn:psi1ofpsitilde}
\psi_1 & =  \Delta^{-1}\big( (\mu M^* N + \mu^*)
\tilde{\psi}_1 - (\mu M^* N - \mu^*) \tilde{\psi}_2 \big),\\
\label{eqn:psi2ofpsitilde}\psi_2 & = \Delta^{-1}( (\mu M^* N - \mu^*)
\tilde{\psi}_1 + (\mu M^* N + \mu^*) \tilde{\psi}_2 \big),
\end{align}
\end{subequations}
where $\Delta \equiv (\mu^*)^2\mathbbm{1}+\mu^2 (M^*N)^2$. For $\Delta$ to
be invertible we must choose $\mu$ such that $\pm i\mu^*/\mu$ is not an
eigenvalue of $M^* N$, which is always possible. Hence, we find for $\psi$
in terms of $\tilde{\psi}$
\begin{equation}
\psi = 2 \Delta^{-1}(\mu M^*N \tilde{\psi}_1 + \mu^* \tilde{\psi}_2).
\end{equation}

We can now show that a reality condition on $\psi$ using
pseudo-conjugation is, in terms of the number of constraints imposed,
equivalent to a reality condition on $\tilde{\psi}$ using standard complex
conjugation. From \refes{eqn:psitilde1ofpsi}{eqn:psi2ofpsitilde} and the
fact that $\psi_1^*=\psi_2^\diamond$ and $\psi_2^*=-\psi_1^\diamond$ we
have
\begin{equation}\label{eqn:psitildereality}
\left. \begin{array}{l}
\psi_1=-N^{-1}\psi_2^* \\
\psi_2=N^{-1}\psi_1^*
  \end{array}\right\}
\iff \left\{\begin{array}{l}
\tilde{\psi}_1 =M^{-1}\tilde{\psi}_2^\diamond  \\
\tilde{\psi}_2= - M^{-1} \tilde{\psi}_1^\diamond
\end{array}\right.
\end{equation}
Now, combining \refec{eqn:standard-symplectic}{eqn:graded-symplectic} with
\refe{eqn:psitildereality} we find that
\begin{equation}
\label{eqn:equality_of_reality_cond} \psi=N^{-1}\psi^\diamond \iff
\tilde{\psi}=M^{-1}\tilde{\psi}^*.
\end{equation}
As there exists an invertible map between $\psi$ and $\tilde{\psi}$, this
proves that a reality condition using pseudo-conjugation imposes the same
number of constraints as does a reality condition using standard complex
conjugation\footnote{Note that in most cases only one of $\psi$ or
$\tilde{\psi}$ can be chosen to have the correct transformation properties
under the Lorentz group in order to be regarded as a spinor. In the cases
where $t - s = 2 \;\mbox{mod}\, 4$, both $\psi$ and $\tilde{\psi}$ can be
chosen to transform as spinors.}.

\subsection{Dirac equation and spinor actions}
If $\eta = +1$, see Table \ref{spinortable}, the Dirac equation for the
corresponding Majorana spinors is not consistent with a mass term
\cite{townsend_kugo}. It will therefore be necessary to distinguish
between the Majorana conditions corresponding to the two possible cases
$\eta=\pm 1$. Consider first the standard Majorana condition. If $\eta=-1$
the spinor will simply be referred to as Majorana ($M$). If however
$\eta=+1$ the spinor will be called \emph{pseudo-Majorana} ($M'$).
Similarly, for the graded Majorana condition, the spinor will be called
graded Majorana ($gM$) if $\eta=-1$ and \emph{pseudo-graded Majorana}
$(gM')$ if $\eta=+1$. See Table \ref{spinortable} for a summary.
Consequently pseudo-Majorana spinors must be massless to be consistent
with the Dirac equation and the same is true for pseudo-graded Majorana
spinors.

Now one should note that the Dirac equation for Majorana spinors cannot
always be derived from an action. Whether or not this is possible depends
on the respective Majorana condition used and on the symmetry properties
of $C\gamma_\mu$ and $C$. The Lagrangian for both standard and graded
Majorana spinors will be of the form
\begin{equation}
\mathcal{L}=\psi^\transpose C(i\gamma^\mu \partial_\mu - m)\psi.
\end{equation}
In the case of standard Majorana spinors one easily finds that for the
action to be non-vanishing one has to require $C\gamma_\mu$ to be
symmetric, and, if massive, we further require the charge conjugation
matrix $C$ to be antisymmetric \cite{nieuwenhuizen}. In the case of graded
Majorana spinors the same conditions apply. Note that in Minkowski
spacetimes we have $(C\gamma_\mu)^\transpose = \epsilon C\gamma_\mu$,
therefore an action involving graded Majorana spinors ($\epsilon=-1$) will
vanish. In Euclidean or other signatures, however, this need not be the
case. In Euclidean signatures, for example, an action involving standard
Majorana spinors is non-vanishing only if $d = 0,1,2 \;\mbox{mod}\, 8$,
whereas an action involving graded Majorana spinors is non-vanishing only
if $d = 2,3,4 \;\mbox{mod}\, 8$.

If instead we consider parity violating Lagrangians of the form
\begin{equation}
\mathcal{L}=\psi^\transpose C\Gamma_5(i\gamma^\mu \partial_\mu - m)\psi
\end{equation}
we require $C\Gamma_5\gamma^\mu$ to be symmetric and, in the case of
massive spinors, we also require $C\Gamma_5$ to be antisymmetric (note
that $d$ must be even for $\Gamma_5$ to exist). Now in Minkowski
spacetimes we have $(C\Gamma_5\gamma^\mu)^\transpose = - \epsilon
(-1)^{d/2}C\Gamma_5\gamma^\mu$. Therefore such an action involving graded
Majorana spinors will be non-vanishing in Minkowski spacetimes only if $d
= 0 \; \mbox{mod} \, 4$, whereas in the case of standard Majorana spinors
we require $d = 2 \;\mbox{mod}\, 4$.

Finally let us consider the Dirac action for a pair of symplectic Majorana
spinors. In this case we have
\begin{equation}
\label{eqn:symplectic_action} \mathcal{L} = \psi^{(i)}{}^\transpose C
\epsilon^{ij} (i\gamma^\mu
\partial_\mu - m)\psi^{(j)}.
\end{equation}
For the action to be non-vanishing we require that $C \gamma^\mu$ be
antisymmetric, and in the massive case we additionally require $C$ to be
symmetric.

\subsection{Standard and graded Majorana--Weyl conditions}
Note that in even dimensions, where we have a choice of matrices
$B_\pm$ for $\eta=\pm1$, it is possible to simultaneously impose the
two corresponding reality conditions. Such spinors will be massless
due to the fact that a pseudo-(graded) Majorana condition has been
imposed. There are four possible cases which we shall analyze
separately.

If $t-s=0 \; \mbox{mod}\, 8$ we can impose both $M$ and $M'$ conditions,
giving
\begin{equation}
\psi=B_-^{-1}\psi^*=B_+^{-1}\psi^*.
\end{equation}
Using \refe{eqn:relationB+B-} we see that a consequence of these two
conditions is that
\begin{equation}\label{eqn:weyl}
\psi=\lambda\Gamma_5\psi .
\end{equation}
This equation will only be consistent if $\lambda=\pm 1$, in which
case \refe{eqn:weyl} is seen to be the Weyl condition for a spinor
with helicity $\lambda$. Note that the Weyl condition can be imposed
on spinors in any even dimensional spacetime. Here, however, the
spinors are also Majorana and we see that consistently imposing both
Majorana conditions has naturally given a \emph{Majorana--Weyl}
($MW$) condition. Note that the helicity of the
resulting Majorana--Weyl spinor depends on the value of $\lambda$
and as such on the relative phase chosen between the matrices $B_+$
and $B_-$ in
\refe{eqn:relationB+B-}\footnote{Remember that the
matrices $B_+$ and $B_-$ are defined up to an overall phase
only.}.
\begin{table}[t!]
\centering {\renewcommand{\arraystretch}{1.3}
\begin{tabular}{|c|c|c|c|c|c|c|c|c|c|c|c|c|}
\hline
\backslashbox{t}{d}& $\:1\:$ & $\:2\:$ & $\:3\:$ & $\:4\:$ & $\:5\:$ & $\:6\:$ & $\:7\:$ & $\:8\:$ & $\:9\:$ & $10$ & $11$ & $12$\\
\hline 0 & $\square$ & \color{grey}$\blacktriangledown$ & $\blacksquare '$
& $\blacklozenge$ & $\blacksquare$ & \color{grey}$\blacktriangle$ &
$\square '$ & $\lozenge$ & $\square$ &
\color{grey}$\blacktriangledown$ & $\blacksquare '$ & $\blacklozenge$ \\
\hline $1$ & $\square '$ & $\lozenge$ & $\square$ &
\color{grey}$\blacktriangledown$ & $\blacksquare '$ & $\blacklozenge$ &
$\blacksquare$ & \color{grey}$\blacktriangle$ & $\square '$ & $\lozenge$ &
$\square$ & \color{grey}$\blacktriangledown$\\ \hline $2$ & &
\color{grey}$\blacktriangle$ & $\square '$ & $\lozenge$ & $\square$ &
\color{grey}$\blacktriangledown$ & $\blacksquare '$ & $\blacklozenge$ &
$\blacksquare$ & \color{grey}$\blacktriangle$ & $\square '$ & $\lozenge$
\\ \hline \hline $t = d$ & $\square '$ & \color{grey}$\blacktriangle$ & $\blacksquare$
& $\blacklozenge$ & $\blacksquare '$ & \color{grey}$\blacktriangledown$ &
$\square$ & $\lozenge$ & $\square '$ & \color{grey}$\blacktriangle$ & $\blacksquare$ & $\blacklozenge$\\
\hline
\end{tabular} \\[2ex]
\begin{tabular}{llll}
$\square$ $M$  \:\:\:\: &  $\square '$ $M'$ \:\:\:\: & $\lozenge$ $MW$
\:\:\:\: & {\color{grey}$\blacktriangle$} $gM$ \& $M'$
\\
$\blacksquare$ $gM$ & $\blacksquare '$ $gM'$ & $\blacklozenge$ $gMW$ &
{\color{grey}$\blacktriangledown$} $M$ \& $gM'$
\end{tabular} } \caption{Possible types of maximal reality conditions in different spacetimes.}
\label{spinortable2}
\end{table}

If $t-s=4 \; \mbox{mod}\, 8$ we can impose both $gM$ and $gM'$ conditions
\begin{equation}
\psi=B_-^{-1}\psi^\diamond=B_+^{-1}\psi^\diamond.
\end{equation}
Again we have as a consequence of these equations that $\psi$ must satisfy
the Weyl condition, \refe{eqn:weyl}, with helicity $\lambda=\pm 1$ for
consistency. We refer to such spinors as \emph{graded Majorana--Weyl}
($gMW$).

If $t-s=2 \; \mbox{mod}\, 8$  we can impose both $gM$ and $M'$ conditions
\begin{equation}
\psi=B_-^{-1}\psi^\diamond=B_+^{-1}\psi^*.
\end{equation}
The Weyl condition, \refe{eqn:weyl}, is no longer satisfied due to the
mixed nature of the Majorana conditions. Instead, a consequence of these
two conditions is
\begin{equation}
\label{eqn:crazy_weyl} \psi=\lambda\Gamma_5\psi^{*\diamond},
\end{equation}
where for consistency we must have $\lambda=\pm i$. Note that, although
$\psi$ is not a true Weyl spinor, if we split $\psi=\psi_1+\psi_2$ as in
\refec{eqn:splitpsi}{eqn:splitpsi2} then the combinations $\psi_1\pm
i\psi_2$ are Weyl. However, the physical interpretation of the condition
in \refe{eqn:crazy_weyl} remains unclear.

If $t-s=6 \; \mbox{mod}\, 8$ we have both $M$ and $gM'$ conditions. This
case is very similar to $t-s=2 \; \mbox{mod}\, 8$.

Table \ref{spinortable2} summarizes which reality conditions may be
imposed in each of the most interesting spacetimes.

\subsection{Four-dimensional Euclidean space}
It is worth mentioning here that when working in even dimensions it
is common to use the Weyl representation for spinors. The Weyl
representation can be defined in full generality
for arbitrary signature in any even dimensional spacetime, however it
is perhaps most familiar in four-dimensional
Minkowski space where the use of two-component spinors with dotted
and undotted indices is quite standard. Here, however, we shall
briefly discuss the case of four-dimensional Euclidean space,
demonstrating how the reality conditions imposed in \cite{hawking,
manin, ivanov} fit into the general scheme of graded Majorana
spinors.

The four-dimensional Euclidean gamma matrices are taken to be
\begin{equation}
\gamma^i=\left(\begin{array}{cc}
0 & -i \sigma_i \\
i \sigma_i & 0
\end{array} \right), \quad
\gamma^4=\left(\begin{array}{cc}
0 & \mathbbm{1} \\
\mathbbm{1} & 0
\end{array}\right), \quad
\Gamma_5=\left(\begin{array}{cc}
\mathbbm{1} & 0 \\
0 & -\mathbbm{1}
\end{array} \right).
\end{equation}
Here $i=1,2,3$ and $\sigma_i$ are the standard Pauli matrices. We
choose the matrices $B_\pm$ in this representation to be
\begin{equation}
B_\pm=\left(\begin{array}{cc}
-\varepsilon & 0 \\
0 & \mp\varepsilon
\end{array}\right),
\end{equation}
where $\varepsilon=i\sigma_2$.
We see from the form of $\Gamma_5$ that the four-component Dirac spinor
decomposes into left- and right-handed two-component spinors, $\phi$ and
$\chi$, as
\begin{equation}
\psi=\left(\begin{array}{c}
\phi \\
\chi
\end{array}\right).
\end{equation}
The graded Majorana conditions, $\psi=B_\pm^{-1}\psi^\diamond$, are then simply
\begin{equation}\label{eqn:gM4de}
\phi=\varepsilon\phi^\diamond, \qquad
\chi=\pm\varepsilon\chi^\diamond.
\end{equation}
Note that with this choice of the matrices $B_\pm$
imposing both graded Majorana conditions implies $\chi=0$ and hence
the resulting spinor will be a left-handed graded Majorana-Weyl
spinor. If we had chosen the opposite relative sign between $B_+$
and $B_-$ the resulting spinor would have been right-handed.

Introducing indices $a,b,\ldots=1,2$ for
left-handed spinors, and $a',b',\ldots=1,2$ for right-handed
spinors, we find for \refe{eqn:gM4de} upon displaying the indices
explicitly
\begin{equation}\label{eqn:gM4deweyl}
\phi_a=\varepsilon_{ab}(\phi_b)^\diamond, \qquad
\chi_{a'}=\pm\varepsilon_{a'b'}(\chi_{b'})^\diamond.
\end{equation}
These expressions may be compared to the reality conditions imposed
in \cite{hawking, manin, ivanov}. Note that in this signature
pseudo-conjugation does not change the index type from primed to
unprimed. This is due to the fact that the left-handed and
right-handed components of Spin(4) do not mix under conjugation
\cite{hawking, ivanov}, a situation which can be
contrasted with, for example, four-dimensional Minkowski space where
conjugation acts to interchange the left-handed and right-handed
components of Spin$(1,3)$ \cite{bluebook}.

\section{Applications to supersymmetry}\label{sec:applications}
\subsection{Real forms of the super Poincar\'e algebra}
We shall now investigate how these new reality conditions can be imposed
to give real forms of super Lie algebras, which will subsequently allow
the derivation of supersymmetric field theories involving graded Majorana
spinors.

Let us define the graded commutator
$[K,L]=KL-(-1)^{\epsilon_K\epsilon_L}LK$, where $\epsilon_{K} = 0$ if $K$
is even and $\epsilon_{K} = 1$ if $K$ is odd (and similarly for $L$). The
generators of the general $\mathcal{N}=1$ super Poincar\'e algebra satisfy
\begin{subequations}
\begin{align}
[M_{\mu\nu},M_{\rho\sigma}] & = \eta_{\mu\sigma}M_{\nu\rho} +
\eta_{\nu\rho}M_{\mu\sigma} - \eta_{\mu\rho}M_{\nu\sigma} - \eta_{\nu\sigma}M_{\mu\rho}, \label{eqn:MMcommutator} \\
[M_{\mu\nu},P_\rho] & = \eta_{\rho\nu}P_\mu -\eta_{\rho\mu}P_\nu,\label{eqn:MPcommutator}\\
[M_{\mu\nu},Q_\alpha] & = - (\sigma_{\mu\nu})_{\alpha}{}^\beta Q_\beta,  \label{eqn:MQcommutator}\\
[Q_\alpha , Q_\beta] & = 2k(\gamma^\mu C^{-1})_{\alpha
\beta}P_\mu\label{eqn:QQcommutator},
\end{align}
\end{subequations}
where all other commutators vanish. Here the even generators $M_{\mu\nu}$
and $P_\mu$, generating rotations and translations, respectively, form the
Poincar\'{e} subalgebra, and $Q_\alpha$ are the odd supersymmetry
generators forming a $2^{\lfloor d/2 \rfloor}$ component spinor. We choose
$(\gamma^\mu)_\alpha{}^\beta$ to correspond to the components of the gamma
matrices and $C^{\alpha\beta}$ to correspond to the components of the
charge conjugation matrix $C$. Note that with these index conventions
$C^{-1} = \left((C^{-1})_{\alpha\beta}\right)$. We have
$\sigma^{\mu\nu}=(1/4)(\gamma^\mu\gamma^\nu-\gamma^\nu\gamma^\mu)$ and $k$
appearing in \refe{eqn:QQcommutator} is a constant phase factor which will
be determined when considering a specific real form of the algebra. Note
that if there is no matrix $C$ available such that $\gamma^\mu C^{-1}$ is
symmetric, see \refe{eqn:gammaCinversetr}, it is not possible to write
down such an $\mathcal{N}=1$ algebra. One may, however, instead consider
an $\mathcal{N} \ge 2$ algebra.

The general element of the super Poincar\'{e} algebra is given by
\begin{equation}
\label{eqn:generalelementalgebra} X = \omega^{\mu \nu} M_{\mu \nu} + x^\mu
P_\mu + \theta^\alpha Q_\alpha.
\end{equation}
Here $\omega^{\mu \nu}$, $x^\mu$ are even supernumbers and $\theta^\alpha$
are odd supernumbers forming a Dirac conjugate spinor. In order to define
a real form of the algebra these coefficients must be constrained by
reality conditions such that the algebra still closes. This can be
achieved by using standard complex conjugation or pseudo-conjugation,
respectively.

To impose reality conditions using pseudo-conjugation we require that
there exists a matrix\footnote{Here we shall assume for
  simplicity that $B$ and
  $C$ are related by $C=B^\transpose A$. In even dimensions there may
  occur more general situations which, using \refe{eqn:relationB+B-},
  can be treated similarly to this case.} $B = (B^{\alpha \beta})$ for which $\epsilon =
-1$. A consistent choice of reality conditions is then given by
\begin{equation}
\label{eqn:realitycond*} (\omega^{\mu \nu})^\diamond = \omega^{\mu \nu},
\quad (x^\mu)^\diamond = x^\mu, \quad (\theta^\alpha)^\diamond B^{\alpha
\beta} = \theta^\beta.
\end{equation}
The condition on $\theta^\alpha$ can be viewed as a graded Majorana
condition imposed on a Dirac conjugate spinor. Note that we consider the
(pseudo-)conjugate of a quantity with an upstairs spinor index to have a
downstairs index, and vice versa. It is easily seen that the Poincar\'{e}
subalgebra of \refec{eqn:MMcommutator}{eqn:MPcommutator} closes under the
reality conditions of \refe{eqn:realitycond*}. To show closure of the full
super Poincar\'{e} algebra let us first consider \refe{eqn:MQcommutator}.
We have
\begin{equation}
[\omega^{\mu \nu}M_{\mu\nu},\theta^\alpha Q_\alpha]  = - \omega^{\mu \nu}
 \theta^\alpha (\sigma_{\mu\nu})_{\alpha}{}^\beta Q_\beta.
\end{equation}
For consistency with \refe{eqn:realitycond*} the coefficient of $Q_\beta$
on the right hand side of the above equation must satisfy
\begin{equation}
- \omega^{\mu \nu} \theta^\alpha (\sigma_{\mu\nu})_{\alpha}{}^\beta = -
\left(\omega^{\mu \nu}\theta^\alpha
(\sigma_{\mu\nu})_{\alpha}{}^\gamma\right)^\diamond B^{\gamma
\beta},
\end{equation}
which is easily checked using the fact that $(\sigma_{\mu\nu})^*=
B\sigma_{\mu\nu}B^{-1}$. Further we see from this that the condition
$(\theta^\alpha)^\diamond B^{\alpha \beta} = \theta^\beta$ is Lorentz
covariant. Finally let us consider \refe{eqn:QQcommutator}. We have
\begin{equation}
[\theta^\alpha Q_\alpha , \tilde{\theta}^\beta Q_\beta]  = - 2k
\theta^\alpha \tilde{\theta}^\beta (\gamma^\mu C^{-1})_{\alpha
\beta}P_\mu.
\end{equation}
For the algebra to close under the reality conditions,
\refe{eqn:realitycond*}, the coefficient of $P_\mu$ on the right hand side
of the equation must be real with respect to pseudo-conjugation. Using
\refe{eqn:gammaCinverse*} we find
\begin{align}
\left(k \theta^\alpha \tilde{\theta}^\beta (\gamma^\mu C^{-1})_{\alpha
\beta}\right)^\diamond &  =  k^* (\theta^\alpha)^\diamond (\tilde
\theta^\beta)^\diamond \left( (\gamma^\mu C^{-1})_{\alpha \beta}
\right)^* \nonumber \\
&  =  k^* \eta^{t + 1}(\theta^\alpha)^\diamond (\tilde
\theta^\beta)^\diamond \left(B \gamma^\mu C^{-1} B^\transpose
\right)^{\alpha \beta} \nonumber \\
&  =  k^* \eta^{t + 1} (\theta^\alpha)^\diamond B^{\alpha \gamma} (\tilde
\theta^\beta)^\diamond (B^\transpose)^{\delta \beta} (\gamma^\mu
C^{-1})_{\gamma \delta}
\nonumber \\
&  =  k^* \eta^{t + 1} \theta^\gamma \tilde \theta^\delta (\gamma^\mu
C^{-1})_{\gamma \delta} \label{eqn:determinek}.
\end{align}
Hence, provided we choose $k$ such that $k= k^*\eta^{t+1}$, the algebra
closes under the reality conditions, \refe{eqn:realitycond*}, which
therefore give a real form of the algebra.

One can alternatively use standard complex conjugation in order to define
a real form of the algebra \refe{eqn:generalelementalgebra}. A consistent
choice of reality conditions on the coefficients is, in this case, given
by
\begin{equation}
\label{eqn:realityconddiamond} (\omega^{\mu \nu})^* = \omega^{\mu \nu},
\quad (x^\mu)^* = x^\mu, \quad (\theta^\alpha)^* B^{\alpha\beta} =
\theta^\beta,
\end{equation}
provided, of course, $B$ is now such that $\epsilon = +1$. That the super
Poincar\'{e} algebra also closes under these conditions can be proven
analogously to the case of pseudo-conjugation. In this case however we
find $k= -k^*\eta^{t+1}$.

In even dimensions we have the possibility of imposing two Majorana
conditions on the coefficients $\theta^\alpha$. Due to the resulting Weyl
condition if $t-s=0,4\; \mbox{mod}\, 8$ we must, in these signatures,
replace \refe{eqn:QQcommutator} with
\begin{equation}\label{eqn:QQcommutatorweyl}
[Q_\alpha,Q_\beta]=2k(\mathbbm{1}+\lambda \Gamma_5)_\alpha{}^\gamma
(\gamma^\mu C^{-1})_{\gamma \beta} P_\mu,
\end{equation}
which is possible provided that both $\Gamma_5\gamma^\mu C^{-1}$ and
$\gamma^\mu C^{-1}$ are symmetric (note that here $C$ is a particular
choice of $C_\pm=B_\pm^\transpose A$). It is then possible to define a
real form of the algebra by imposing $MW$ or $gMW$ conditions on the Dirac
conjugate spinor $(\theta^\alpha)$ with corresponding reality conditions
on the $\omega^{\mu \nu}$'s and $x^\mu$'s. For example, let us consider
$t-s=4\; \mbox{mod}\, 8$. The algebra will close if we impose the $gMW$
condition
\begin{equation}
\theta^\alpha= (\theta^\beta)^\diamond (B_-)^{\beta \alpha}=
(\theta^\beta)^\diamond(B_+)^{\beta \alpha}
\end{equation}
along with the conditions $(\omega^{\mu \nu})^\diamond = \omega^{\mu \nu}$
and $(x^\mu)^\diamond = x^\mu$. If $t-s=2,6\; \mbox{mod}\, 8$ we may
consistently impose both a graded and a standard Majorana condition on the
coefficients $\theta^\alpha$. However, the physical interpretation of such
mixed reality conditions remains unclear.

\subsection{Three-dimensional Euclidean field theory}
In order to illustrate the applications of graded Majorana spinors to
supersymmetric field theories let us construct a simple example in
three-dimensional Euclidean space (i.e., $t=3$, $s=0$). From Table
\ref{spinortable} we see that $\epsilon=-1$ and so no standard Majorana
spinors exist. We choose the gamma matrices to be the standard Pauli
matrices $\gamma^i=\sigma_i = \left((\sigma_i)_\alpha{}^\beta\right)$,
$i=1,2,3$, and we take $B=\varepsilon = (\varepsilon^{\alpha \beta})$.
Here $\alpha = -,+$ are two-spinor indices and the quantity
$\varepsilon^{\alpha\beta}$ is the invariant antisymmetric tensor with
$\varepsilon^{-+}=+1$.  We use $\varepsilon^{\alpha\beta}$ to raise
indices, with the convention
$\psi^\alpha=\varepsilon^{\alpha\beta}\psi_\beta$. Indices will be lowered
using $\varepsilon_{\alpha\beta}$, $\varepsilon_{-+}=+1$, with the
convention $\psi_\alpha=\psi^\beta\varepsilon_{\beta\alpha}$. If we define
$J_i=-\frac{1}{2}\epsilon_{ijk}M_{jk}$, then the $\mathcal{N}=1$ super
Poincar\'e algebra can be rewritten as
\begin{subequations}
\begin{align}
[J_i,J_j] &= \epsilon_{ijk}J_k, \\
[J_i,P_j] &= \epsilon_{ijk}P_k, \\
[J_i,Q_\alpha] &= \frac{i}{2}(\sigma_i)_\alpha{}^\beta Q_\beta, \\
[Q_\alpha,Q_\beta] &= 2i(\sigma_i\varepsilon)_{\alpha\beta}P_i.
\end{align}
\end{subequations}
Writing the general element of the algebra as $X=\varphi^i J_i+ x^i P_i +
\theta^\alpha Q_\alpha$ we obtain a real form by imposing reality
conditions $(\varphi^i)^\diamond=\varphi^i$, $(x^i)^\diamond=x^i$ and
$(\theta^\alpha)^\diamond B^{\alpha\beta}= \theta^\beta$. Exponentiating
the algebra gives the super Poincar\'e group, $S\Pi$, from which we form
the coset space $S\Pi / SO(3)$, where $SO(3)$ is the rotation group
generated by the $J_i$. Following the method discussed in \cite{west} we
consider a coset representative
\begin{equation}
L(x^i,\theta^\alpha)=\mbox{exp}(x^iP_i+\theta^\alpha Q_\alpha),
\end{equation}
so that $(x^i,\theta^\alpha)$ are coordinates on the coset space. We hence
have $S\Pi/SO(3)=\R^{3|2}$ where reality is defined with respect to
pseudo-conjugation as given above.

The left action of $S\Pi$ on the coset representative induces a
transformation on the coordinates $(x^i,\theta^\alpha)\to(x^i+\delta x^i
,\theta^\alpha+ \delta \theta^\alpha)$. Using this we can find the
differential operator representation of the generators of the
superalgebra. In particular we have,
\begin{equation}
Q_\alpha=-\partial_\alpha+i(\sigma_i\varepsilon)_{\alpha\beta}\theta^\beta
\partial_i.
\end{equation}

An invariant vielbein $(E^i, E^\alpha)$ and spin-connection $\Omega^i$ on
$\R^{3|2}$ can be constructed from the coset representative as
\begin{equation}
L^{-1}dL=E^iP_i+E^\alpha Q_\alpha +\Omega^i J_i.
\end{equation}
We find that $\Omega^i=0$, and so the inverse vielbein determines the
covariant derivatives, which turn out to be
\begin{subequations}
\begin{align}
D_i &= \partial_i, \\
D_\alpha &=
\partial_\alpha+i(\sigma_i\varepsilon)_{\alpha\beta}\theta^\beta\partial_i.
\end{align}
\end{subequations}

For an even superscalar field $\Phi(x,\theta)$, satisfying
$\Phi^\diamond=\Phi$, let us consider the action
\begin{equation}
I=\int d^3xd\theta^-d\theta^+ \left(\frac{1}{2}D^\alpha\Phi D_\alpha\Phi-
U(\Phi)\right).
\end{equation}
It is easily seen that $[Q_\alpha, D_\beta]=0$, from which it follows that
this action will be invariant under supersymmetry transformations $\delta
\Phi=\beta^\alpha Q_\alpha \Phi$. We can expand $\Phi$ in component fields
as
\begin{equation}
\Phi(x,\theta)=A(x)+\theta^\alpha
\psi_\alpha(x)+\frac{1}{2}\theta^\alpha\theta_\alpha F(x).
\end{equation}
The condition $\Phi^\diamond=\Phi$ yields $A=A^\diamond$, $F=F^\diamond$
and $\psi_\alpha=(B^{-1})_{\alpha\beta}(\psi_\beta)^\diamond$. Hence we
see that $\psi$ is a graded Majorana spinor.

The action $I$ can be rewritten in terms of the component fields. Upon
elimination of the auxiliary field $F$ via its equations of motion, and
integrating out the $\theta$ coordinates, $I$ becomes
\begin{equation}
I = \int d^3x \Big( (\partial A)^2 -\frac{1}{4} U'(A)^2 + i
\psi^\alpha(\sigma^i)_\alpha{}^\beta \partial_i \psi_\beta
+\frac{1}{2}U''(A)\psi^\alpha\psi_\alpha \Big).
\end{equation}
This is the action for a real scalar field coupled to a graded Majorana
spinor in three-dimensional Euclidean space. For an example of a
supersymmetric action involving Dirac spinors in this signature see
\cite{mckeon}. Note that, as $C\gamma^\mu$ is symmetric in this signature,
a supersymmetric action containing a symplectic action of the form of
\refe{eqn:symplectic_action} does not exist.

\section{Conclusions and Outlook}
We have seen how the classification of possible reality conditions on
Grassmann odd valued spinors should be extended by what we call a graded
Majorana condition. In contrast to the symplectic Majorana condition
which, in order to be imposed, requires an even number of spinor fields,
the graded Majorana condition can be imposed on a single spinor. In fact,
as we showed in Section \ref{sec:equivalence_reality_conditions} the
graded Majorana condition imposes the same number of constraints on a
spinor as does a standard Majorana condition.

In order to illustrate the use of graded Majorana spinors in
supersymmetric field theories we constructed an action involving such
spinors in the case of three-dimensional Euclidean space. In globally
curved space an example of the use of graded Majorana spinors is obtained
by considering field theories on the supersphere
$S^{2|2}=\uosp/\mathcal{U}(1)$, as investigated in \cite{supersphere}.
Graded Majorana spinors could also play an important role in the
construction of supergravity theories.  In this context, an interesting
example of a spacetime where no standard Majorana spinors exist is
11-dimensional Euclidean space. It will be very interesting to investigate
whether the existence of graded Majorana spinors may account for a
physically sensible supergravity theory in this spacetime.

\section*{\normalsize{Acknowledgements}}
\vspace{-2ex} The authors would like to thank Professor N.\/~S.\/~Manton
for many helpful conversations.

This work was partly supported by the UK Engineering and Physical Sciences
Research Council. One of the authors (A.F.K.) gratefully acknowledges
financial support by the Gates Cambridge Trust.

\providecommand{\href}[2]{#2}\begingroup\raggedright\endgroup

\end{document}